# From Automated Simulation to Autonomous Discovery: A Hierarchical Framework for Agentic Computational Materials Science

Linggang Zhu, Jian Zhou, Zhimei Sun*

School of Materials Science and Engineering, Beihang University, Beijing 100191, China.

*Corresponding author: zmsun@buaa.edu.cn

**Abstract**

The convergence of large language models, materials-specific foundation models, and agentic artificial intelligence is reshaping the paradigm of computational materials discovery. While high-throughput computation, automated workflows, and data-driven modeling have greatly expanded the scale of materials exploration, the core scientific decision-making loop remains largely human-directed. Agentic AI introduces the possibility of systems that can autonomously reason about materials objectives, execute simulations, and refine strategies. However, the rapid emergence of such systems has created a critical need for a unified and operational framework to define, evaluate, and guide scientific autonomy in computational materials discovery. In this Perspective, we propose the Computational Materials Agent Autonomy Level (CMA-AL) framework, a hierarchical taxonomy defining six levels of autonomous agency in computational materials science: scripted excecutor, LLM-assisted operator, adaptive explorer, experiment-ready modeler, agentic digital twin, and self-extending intelligence. We further map emerging agentic systems onto the framework and identify key scientific and technological challenges toward higher autonomy. CMA-AL provides a common language for characterizing agentic computational materials discovery, evaluating the maturity of emerging systems, and guiding their evolution toward increasingly autonomous materials discovery.

## 1. Introduction

The discovery of new materials has historically relied on the integration of diverse scientific disciplines, including quantum mechanics, thermodynamics, kinetics, materials synthesis, and processing. This intrinsic complexity makes materials discovery both scientifically rich and cost challenging. Computational materials science has fundamentally transformed this landscape by enabling researchers to perform increasingly sophisticated in silico experiments, reducing dependence on conventional trial-and-error approaches.[1,2] Over the past two decades, advances in simulations across multiple spatial and temporal scales, together with high-throughput computational frameworks and large-scale materials databases, have enabled systematic exploration of vast chemical spaces.[3-9]

Despite these advances, the dominant paradigm of computational materials discovery remains fundamentally human-directed. Although automated workflows can execute thousands or even millions of calculations with minimal manual intervention, the scientific decision-making process itself remains largely controlled by researchers. Humans define the design objective, select appropriate computational methods, determine simulation parameters, interpret intermediate results, assess physical plausibility, and decide the next computational or experimental step. Automation has accelerated individual operations, but it has not fundamentally changed the structure of scientific reasoning. The next frontier of computational materials discovery is therefore not simply increased computational throughput, but autonomous scientific agency. The rapid emergence of large language models (LLMs), tool-augmented reasoning systems, and multi-agent architectures has created unprecedented opportunities toward the goal of autonomous science.[10-18]

Literally, autonomy refers to "the capacity to think, decide, and act on the basis of such thought and decisions freely and independently and without let or hindrance".[19] This conceptualization provides a useful starting point for considering autonomy in computational materials agents. In this context, autonomy can be operationalized according to the extent to which an agent can translate high-level materials design goals into executable computational workflows, assess intermediate outcomes, refine its strategies through feedback, and ultimately derive scientifically meaningful knowledge with reduced human intervention. As this emerging research direction begins to develop, the community faces a critical challenge: there is currently no unified and operational framework to define, evaluate, and guide the evolution of autonomous capabilities in computational materials discovery. Existing efforts remain highly fragmented, with many systems focusing on isolated capabilities, while similar functionalities are often independently developed[20-22] and the pathway from workflow automation toward fully autonomous materials discovery remains unclear. Therefore, a roadmap for autonomous computational materials discovery is urgently needed, not only to accelerate the development of agentic systems but also to establish a common framework for defining, evaluating, and advancing computation autonomy. Such a roadmap should provide a systematic description of how autonomous capabilities evolve and identify the key milestones toward fully autonomous materials discovery. To achieve this goal, several fundamental questions need to be addressed: (1) What scope of the materials discovery workflow can an agent autonomously control? This includes the transition from individual calculations toward complete design-simulation-validation loops; (2) What computational capabilities are required to achieve different levels of autonomy? These range from executing established workflows to integrating multi-scale physical models and eventually developing new computational methodologies; (3) What degree of human involvement and scientific oversight is required at each autonomy level? This concerns maintaining scientific accountability, interpretability, and reliability as autonomous capabilities increase.

## 2. CMA-AL framework defining the autonomy level of computational materials agent

### 2.1 Overview

Here, we propose the Computational Materials Agent Autonomy Level (CMA-AL) framework that defines six levels of autonomous agency (L0-L5) in computational materials discovery. The CMA-AL framework is intended as an operational classification of autonomy rather than an anthropomorphic categorization of agents as assistants, co-scientists, or scientists. The framework is organized along two complementary dimensions. The first dimension represents computation autonomy, describing an agent's capability to orchestrate computational workflows, integrate heterogeneous tools and models, and ultimately extend existing computational frameworks. The second dimension represents human responsibility, capturing how the role of human researchers evolves across different levels of autonomy. At lower levels, humans remain responsible for workflow design, method selection, and result interpretation. With increasing autonomy, human roles transition from direct operational control toward defining research objectives, supervising scientific decisions, and ensuring responsible governance of autonomous discovery processes. The overall framework is illustrated in Figure 1, with its key components and details discussed in the following sections. Notably, the levels are cumulative: higher levels of autonomy retain and build upon the capabilities established at lower levels while introducing additional forms of scientific reasoning, coordination, and decision-making.

### 2.2 Level definitions and boundary in-between

**L0: Scripted executor**

At L0, the computation capability is limited to the automated execution of predefined computational workflows. The agent functions as a computational executor that removes repetitive manual operations but does not participate in scientific decision-making. Typical L0 systems integrate existing simulation software with workflow engines, enabling automated structure preparation, job submission, error handling based on predefined rules, data parsing, and database construction. However, the workflow logic itself is determined before execution. The human researcher defines the scientific objective, selects the computational method, determines simulation parameters, designs the workflow structure, and interprets the final results. Intermediate outcomes generally do not modify the scientific strategy unless explicitly encoded by the workflow designer. L0 systems provide the essential computational infrastructure for higher-level agents, but they do not themselves represent autonomous scientific intelligence.

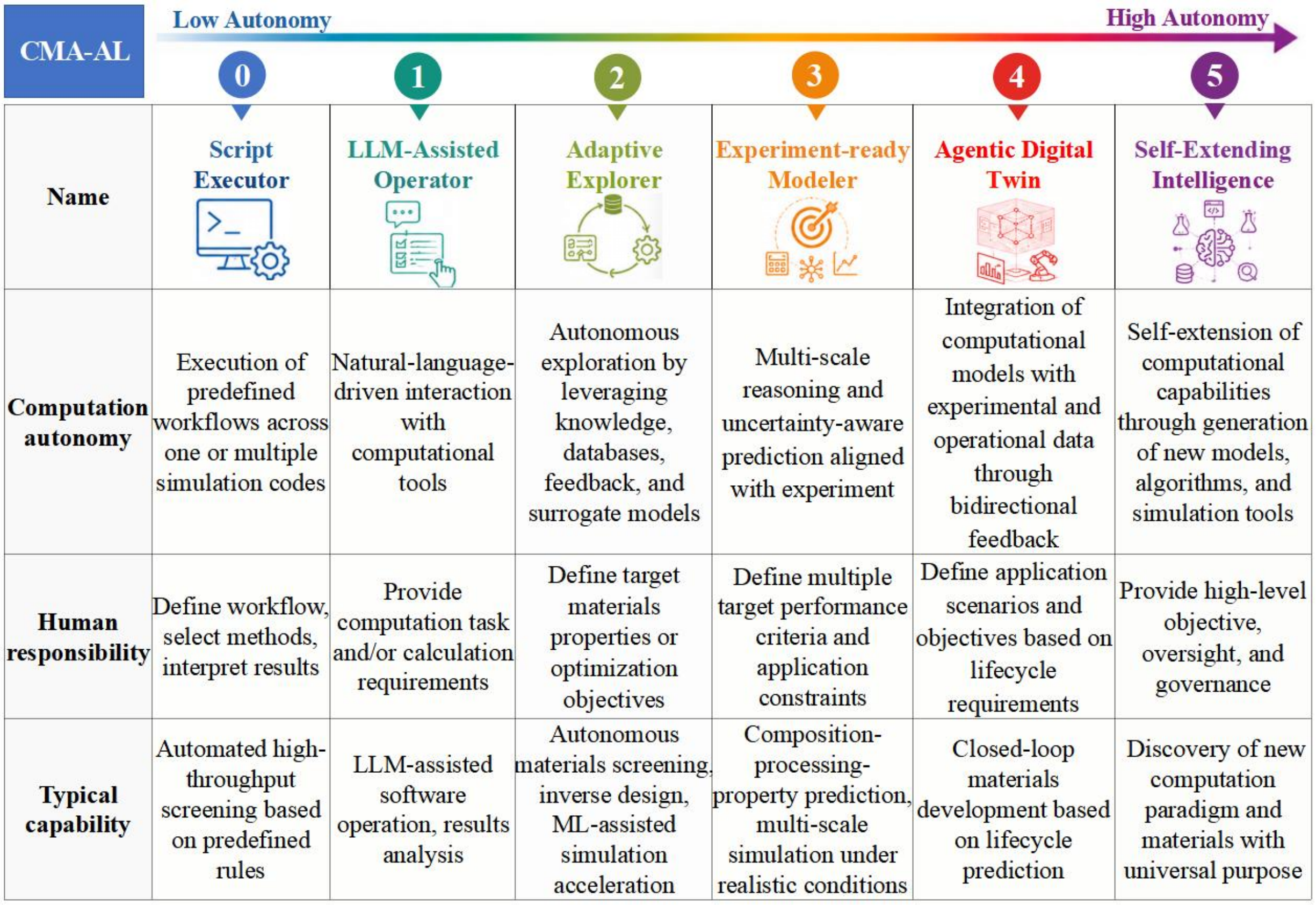

| CMA-AL (Low Autonomy → High Autonomy) | 0 | 1 | 2 | 3 | 4 | 5 |
|---|---|---|---|---|---|---|
| **Name** | Script Executor | LLM-Assisted Operator | Adaptive Explorer | Experiment-ready Modeler | Agentic Digital Twin | Self-Extending Intelligence |
| **Computation autonomy** | Execution of predefined workflows across one or multiple simulation codes | Natural-language-driven interaction with computational tools | Autonomous exploration by leveraging knowledge, databases, feedback, and surrogate models | Multi-scale reasoning and uncertainty-aware prediction aligned with experiment | Integration of computational models with experimental and operational data through bidirectional feedback | Self-extension of computational capabilities through generation of new models, algorithms, and simulation tools |
| **Human responsibility** | Define workflow, select methods, interpret results | Provide computation task and/or calculation requirements | Define target materials properties or optimization objectives | Define multiple target performance criteria and application constraints | Define application scenarios and objectives based on lifecycle requirements | Provide high-level objective, oversight, and governance |
| **Typical capability** | Automated high-throughput screening based on predefined rules | LLM-assisted software operation, results analysis | Autonomous materials screening, inverse design, ML-assisted simulation acceleration | Composition-processing-property prediction, multi-scale simulation under realistic conditions | Closed-loop materials development based on lifecycle prediction | Discovery of new computation paradigm and materials with universal purpose |

**Figure 1.** Computational Materials Agent Autonomy Level (CMA-AL) framework. Six levels of autonomy (L0-L5) are defined along two dimensions: computation autonomy and human responsibility. The representative capabilities and evolutionary progression of computational

materials agents at each level are also illustrated.

**L1: LLM-assisted Operator**

At L1, large language models introduce a semantic interface between human researchers and computational software, while providing limited reasoning capabilities for well-defined scientific tasks. Instead of manually constructing input files and executing software-specific commands, users can describe computational objectives through natural language, and the agent translates these intentions into executable simulation workflows. For specific computation software, an L1 agent can generate input files, retrieve software-specific knowledge, configure calculations based on established guidelines, submit jobs, monitor execution status, recover from known technical errors, and summarize computational results. Achieving reliable L1 capability requires domain-specific resources, including knowledge bases or skill libraries, to provide software documentation, input conventions, parameter recommendations, and predefined error-handling strategies.

**L1/L0 boundary: From scripted execution to semantic translation.** The L1/L0 boundary represents the transition from rigid, hardcoded computational scripts to dynamic, natural-language-driven calculation orchestration. While an L0 agent executes static workflows where every logical branch, input parameter, and file parser must be hardcoded at design time by a human programmer, an L1 agent interprets high-level human instructions and dynamically generate computational procedures by mapping scientific descriptions to software operations. Crossing this boundary eliminates the requirement for human researchers to write custom shell scripts or manually format software-specific input files.

**L2: Adaptive explorer**

At L2, the agent gains the capability to autonomously adjust computational workflows based on accumulated knowledge and feedback from previous observations. The agent can integrate information from scientific literature, materials databases, prior calculations, intermediate simulation results, or experimental data to optimize computational parameters, refine workflow decisions, and determine subsequent actions within an established computational framework. This adaptive capability enables efficient exploration of large and complex materials spaces that are impractical to investigate through conventional sequential high-throughput calculations. Active-learning-based materials discovery represents a typical L2 capability. Similarly, machine-learning interatomic potential (MLIP) workflows can reach L2 autonomy when the agent autonomously determines training data selection, model refinement. L2 autonomy also enables adaptive inverse design, where agents iteratively generate and refine candidate materials toward single or multiple target properties using approaches such as genetic algorithms or generative machine-learning models and subsequently refine the search based on computational evaluations.

**L2/L1 boundary: From assisted execution to autonomous exploration.** An L1 agent can autonomously generate and execute a sequential workflow within a specific software environment following a human-defined objective or calculation trajectory. In contrast, an L2 agent can establish an autonomous feedback loop: objective-simulation-evaluation-next decision. The defining shift is the introduction of a closed-loop feedback mechanism where computed material properties are not merely returned to the user, but are immediately ingested as active data points to refine subsequent calculations.

**L3: Experiment-ready modeler**

At L3, the agent becomes a computational architect capable of constructing and managing complex simulation pipelines across different physical scales and softwares. Materials discovery rarely relies on a single property calculation; realistic materials design requires the integration of complementary computational approaches that describe different aspects of materials behavior. An L3 agent can determine which computational methods are required for a given scientific objective and dynamically construct a coherent multi-method workflow. For example, designing a new material may involve DFT calculations for bonding and defect energetics, machine-learning potentials for long-time dynamic, CALPHAD and phase-field models for phase stability and microstructure evolution, and continuum approaches for processing and mechanical performance. A key capability emerging at L3 is experiment-oriented computational prediction. The agent moves beyond predicting isolated material properties and begins to evaluate practical material viability by considering synthesis constraints, processing conditions, operating environments, and uncertainty associated with computational predictions.

**L3/L2 boundary: From adaptive exploration to methodological orchestration.** An L2 agent operates within a predefined modeling framework and adaptively explores the design space by refining parameters, selecting candidates, or acquiring new data. In contrast, an L3 agent determines how a materials problem should be computationally formulated by selecting, integrating, and coordinating appropriate models and simulation methods across physical scales and software platforms. In short, L2 agents improve existing computational processes, whereas L3 agents construct application-oriented computational pathways that bridge intrinsic property prediction and experimentally relevant materials assessment.

**L4: Agentic digital twin**

At L4, autonomous computational materials discovery extends beyond the virtual environment by establishing continuous interaction between computational models and physical materials systems. The concept of the agentic digital twin here extends the previous concept proposed by Timms et al[23] and Burr et al[24], by integrating emerging digital twin frameworks for materials[25] and the autonomous computational capabilities of L3 agents. L4 agent can incorporate experimental measurements, update computational models, adapt simulation strategies, recommend synthesis and processing conditions, and predict material evolution under realistic operating environments. By interfacing with automated synthesis platforms, characterization systems, and manufacturing environments, the agent provides the computational intelligence layer required for closed-loop materials innovation.

Compared with traditional digital twins that primarily rely on empirical sensing and data assimilation, L4 agents combine experimental feedback with multi-scale modeling to provide mechanistic understanding and high spatial-temporal resolution predictions beyond direct measurements. They can also compensate for mismatches between experimental and simulation scales by providing computational predictions when direct observation is impractical. Importantly, L4 experiment-simulation interaction extends beyond property validation or candidate ranking as achieved at L2, and can incorporate feedback from fabrication and processing stage.

**L4/L3 boundary: From digital model to digital twin.** The transition from L3 to L4 represents the transition from purely computational autonomy to physical-world scientific autonomy. An L3 agent can solve complex materials problems by selecting and orchestrating multiple computational methods, but its operation remains confined to the digital domain. The workflow begins and ends with computational models, and interactions with experiments require human coordination. An L4 agent establishes a persistent computational-experimental loop by directly interacting with physical systems. The defining shift is the transition from a computational assistant that designs simulation strategies to an autonomous scientific system that formulates hypotheses, tests predictions through experiments, and continuously updates its understanding based on real-world observations.

**L5: Self-extending intelligence**

At L5, the agent reaches the highest level of autonomy by acquiring the capability to extend its own computational foundations. Unlike lower-level agents that rely on existing simulation software, computational models, or analysis procedures, which are all pre-developed by human, an L5 agent can identify fundamental limitations in available methodologies when facing new objective and autonomously develop new computational approaches to overcome them. This self-extension capability includes constructing improved physical representations, generating new simulation algorithms, developing advanced analysis methods, and integrating newly developed capabilities into existing computational workflows. Importantly, self-extension is not simply triggered by the absence of a suitable software tool; rather, it emerges when the agent recognizes fundamental limitations of existing computational paradigms, such as systematic discrepancies between predictions and reliable experimental observations, incomplete physical descriptions, or unresolved scientific phenomena. Upon reaching L5 autonomy, the complete framework and its key components, encompassing the full hierarchy from L0 to L5, are illustrated in Figure 2.

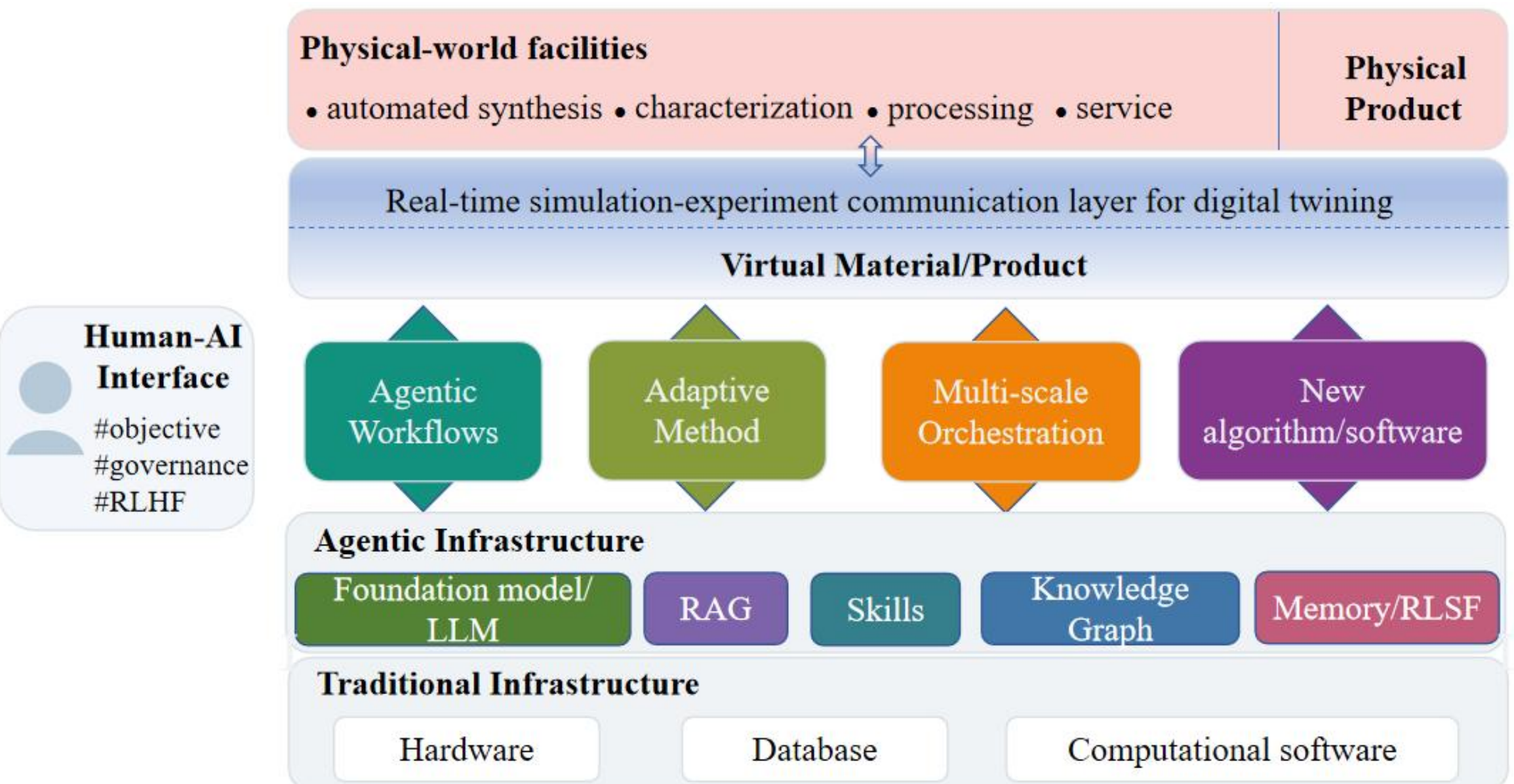

**Figure 2.** Schematic illustration of the integrated agentic computational materials framework, encompassing autonomy levels L0–L5. The foundation model provides core capabilities in reasoning and code generation. Abbreviations: RLHF, reinforcement learning from human feedback; RLSF, reinforcement learning from scientific feedback.

**L5/L4 boundary: From autonomous operation to self-extension.** The transition from L4

to L5 represents the difference between an autonomous user of scientific methodologies and an autonomous developer of new scientific capabilities. An L4 agent can independently conduct complex materials research by selecting, integrating, and adapting existing computational models, simulation methods, and experiment-simulation feedback loops. In contrast, an L5 agent can recognize fundamental limitations of existing computational approaches and extend the methodological foundations of scientific discovery itself. This represents the ultimate vision of agentic computational materials discovery: not merely accelerating research within existing computational paradigms, but continuously expanding the boundaries of how materials phenomena can be represented, simulated, and understood.

## 3. Current Landscape of agentic Systems Across Autonomy Levels

Existing agentic systems span a range of autonomy levels within CMA-AL, with most current developments concentrated at L0-L2 and only a limited number demonstrating capabilities approaching L3. L4 and L5 autonomy remain largely at the conceptual stage. The details are discussed in the following sections.

### 3.1 Agents at L0-L2

L0 systems include pymatgen[4], AiiDA[26], Atomate[27], AFLOW[28], ALKEMIE[7] and JAMIP[8] that have established infrastructures for automated calculation execution, data management, and reproducible computational workflows. These platforms enable high-throughput materials exploration by automatically performing repetitive computational tasks.

At L1, VASPilot[29] integrates LLM-based agents with software tools through the Model Context Protocol (MCP), enabling structure manipulation, input generation, job submission, and post-processing for VASP calculations. LAMMPS-agent[20] combines multiple open-source tools to automate molecular dynamics workflows, while CrystalPlasticitySim[30] employs multiple agents for input generation, simulation execution, result extraction, and parameter optimization, involved in crystal plasticity simulations.

A growing number of systems demonstrate elements of L2 adaptive computation by using intermediate results, prior knowledge, surrogate models or on-the-fly experimental feedback to guide subsequent computational decisions. AtomAgents leverages the high reasoning capabilities of the all-purpose LLM GPT-4 family models, incorparate a research-hypothesis generation module for efficient simulation.[21] MASTER uses hierarchical reasoning and active learning to select subsequent DFT calculations for catalyst-design problems.[31] LLMatDesign employs LLM agents to modify structures, evaluate outcomes, and refine decisions based on previous results. By incorporating self-reflection mechanisms, the system can adapt computational strategies to different design objectives.[32] MAPPS adds goal-conditioned workflow planning, generated tool code, physics-based evaluators, scientist feedback, and error reflection.[33] MatPC integrates LLM reasoning with crystal structure prediction, similarity evaluation, dimensionality reduction, and DFT validation.[34] Another multi-agent orchestration framework demonstrated at Argonne National Laboratory's Aurora supercomputer[35], which deploys a planner-executor hierarchy to screen MOF candidates across DFT and force-field codes, enables efficient and scalable execution on the Aurora supercomputer. The Autonomous Materials Search Engine (AMASE) enables efficient mapping of temperature-composition phase diagrams through a self-driving loop that continuously integrates experimental measurements with computational predictions.[36] Together, these systems indicate that adaptive computational exploration is emerging, although the extent of autonomous decision-making varies substantially among implementations.

### 3.2 Agents approaching L3

Several emerging systems are beginning to approach L3 by integrating multiple computational capabilities into broader research environments designed to address realistic materials design problems. ALKEMIE Agent[37] represents such an integrated framework by combining materials recommendation, knowledge retrieval, autonomously generated simulation workflows, and active-learning strategies for materials screening. In ALKEMIE Agent, multiscale simulation workflows, including DFT-Monte Carlo (DFT-MC) and universal machine-learning interatomic potential-Monte Carlo (uMLIP-MC) simulations, enable the exploration of atomic configurations and phase distributions in complex multicomponent systems. MatSciAgent similarly routes natural-language requests to materials databases, crystal-generation utilities, molecular-dynamics tools, and continuum simulations, although its demonstrated tasks remain bounded.[38] Similar multi-fidelity simulation workflows have also been adopted in other emerging materials agents, such as Matty[39] and El Agente Q[40] in materials chemistry.

However, these systems have not yet fully achieved L3 autonomy as defined in this framework. Remaining challenges include reliable orchestration across heterogeneous computational methods, uncertainty quantification and propagation throughout multi-stage workflows, and the ability to translate computational predictions into experimentally relevant materials design strategies.

### 3.3 Foundations toward L4 and L5

L4-L5 levels remain largely aspirational, although several emerging concepts provide preliminary foundations. Digital twin concepts in chemistry and materials science have demonstrated the potential of continuously updated computational representations of physical systems. For example, digital twins for chemical science (DTCS)[41] aim to integrate simulation and characterization data to provide dynamic understanding of molecular and material processes, however, in DTCS the expensive first-principles calculations are performed off-site, stored as intrinsic property of the species for following prediction. Most existing digital twin frameworks for materials remain simulation-informed rather than truly agentic. They can maintain updated representations of physical systems, but they generally lack autonomous close-loop interaction between computation and experiment. Finally, self-extending L5 agents require capabilities beyond current AI systems, and no verified computational materials agent currently reaches L5. Symbolic regression, neural operators, and automated program search can generate equations or algorithms under human-defined representations and evaluators, but L5 requires a longer chain: recognizing that existing physics or numeric is inadequate; proposing a genuinely new representation or method; implementing and validating it against independent simulations; and safely incorporating it into future workflows. Emerging concepts such as recursive self-improvement (RSI) represent potential future directions toward this level of autonomy.

## 4. Future Challenges and Outlook

While recent agentic systems have demonstrated remarkable progress in workflow automation, natural-language interaction, and computational optimization, achieving higher levels of autonomy requires addressing fundamental challenges at the intersection of Agentic AI and computational materials science. Here, we identify five key challenges that will determine the future development of autonomous computational materials discovery.

### 4.1 Scientific reasoning, hypothesis generation, and materials pre-screening before computation

General-purpose LLMs have demonstrated remarkable reasoning capabilities, but hallucinations remain a persistent challenge.[42] Meanwhile, autonomous computational materials discovery requires a deeper form of scientific reasoning. An effective materials agent must understand not only how to execute computational tools, but also why a specific scientific question, material system, structure, or simulation pathway is meaningful. Future agents must integrate physical principles, existing scientific knowledge, and computational experience to identify governing mechanisms, generate promising hypotheses, define meaningful design spaces, evaluate feasibility, and prioritize candidates. Without such capabilities, agents may produce computationally valid but scientifically redundant, physically implausible, or experimentally irrelevant candidates.

From a technical perspective, several enabling components for such agents are already emerging. Materials knowledge graphs and ontologies can connect compositions, structures, properties, synthesis operations, characterization methods, and provenance across databases and the literature. MatKG, for example, demonstrates literature-scale extraction into a graph representation[43], while Propnet shows how explicit property relations can propagate knowledge and expose derivation paths.[44] Literature-derived concept graphs and link prediction can rank unusual concept combinations as prompts for expert ideation.[45] SciAgents has explored a knowledge-graph-guided multi-agent framework for generating and iteratively refining materials hypotheses.[46] Collectively, these approaches can expand the accessible scientific search space while providing more transparent and inspectable evidence pathways. Nevertheless, important limitations remain: graph edges may reflect statistical association or co-occurrence rather than causal relationships, and knowledge extracted from the literature inevitably inherits biases, incompleteness, and inconsistencies present in the published record.

**4.2 Reliability and uncertainty in autonomous computational workflows**

A second major challenge is ensuring reliability as agentic workflows become increasingly complex. Unlike conventional simulation pipelines, in which researchers manually inspect critical decisions and intermediate results, autonomous materials agents involve interconnected processes spanning scientific reasoning, knowledge retrieval, workflow planning, tool and method selection, numerical simulation, machine-learning prediction, and result interpretation. Uncertainty in such systems therefore cannot be represented by a single numerical error bar, but instead constitutes a heterogeneous and propagated form of uncertainty arising from the agent itself, the underlying data and machine-learning models, and the physical and numerical approximations of materials simulations.

Recent work on uncertainty quantification (UQ) for LLM agents similarly emphasizes that agentic systems introduce uncertainty over heterogeneous entities and, importantly, dynamically evolving uncertainty across interactive decision processes.[47] At the computational level, uncertainty mainly originates from physical simulation and machine-learning model training. A range of established UQ approaches can provide part of the solution. In atomistic machine learning, ensemble disagreement and related uncertainty indicators have already been extensively used to identify configurations that are insufficiently represented by the training data and to trigger additional high-fidelity calculations.[48,49] Uncertainty also exists within first-principles calculations themselves: for example, Bayesian error estimation has been used to quantify uncertainty associated with exchange-correlation approximations in DFT and to propagate this uncertainty into derived quantities such as phonon properties and lattice thermal conductivity.[50,51]

The more difficult problem, however, is workflow-level uncertainty propagation. In an autonomous multiscale workflow, uncertainty introduced at one stage may alter both the input and the methodological choice of subsequent stages. Errors or uncertainty in first-principles data can propagate into an MLIP, affect molecular-dynamics trajectories, and ultimately alter predicted kinetic or thermodynamic properties. Such cross-model and cross-scale uncertainty propagation has long been recognized as a challenge in integrated computational materials engineering (ICME).[52] Methods developed for multi-fidelity uncertainty propagation and model fusion provide an important foundation for addressing such coupled uncertainties, particularly when computational models of different fidelity and cost are combined.[53] Ultimately, achieving reliable autonomous materials discovery will require moving beyond uncertainty quantification of individual models toward workflow-level uncertainty management, in which uncertainty becomes an explicit variable governing the agent's actions.

### 4.3 Physics-grounded simulation-experiment fusion

The transition toward L4 autonomy requires overcoming the separation between computational prediction and experimental reality. Simulations provide atomic-scale insights into material mechanisms but inevitably involve approximations, including limited system sizes, accessible time scales, and imperfect theoretical models. Experiments, in contrast, directly characterize real materials but often provide incomplete information about underlying mechanisms. Bridging these complementary descriptions therefore requires more than simply connecting computational and experimental workflows; it requires a physics-grounded, bidirectional framework in which simulations guide experimental design, while experimental observations update the computational models and redirect subsequent simulations and decisions.

Achieving this goal requires computational frameworks that balance accuracy, efficiency, and scalability, particularly for chemically complex, multi-component, and multi-scale materials systems. High-fidelity methods alone are often insufficient for continuous autonomous operation due to their computational cost. Therefore, for computationally demanding tasks, L4 and beyond agents will require adaptive computational strategies, including reduced-order models, surrogate models, multi-fidelity simulations, and uncertainty-aware acceleration methods. Multi-fidelity Bayesian optimization already provides a framework for combining information sources of different accuracy and cost in materials discovery.[54] Emerging universal machine-learning interatomic potentials, including CHGNet,[55] MACE,[56] DPA,[57] and NEP89,[58] represent another promising direction by providing a balance between predictive accuracy and computational efficiency for large-scale simulations. It is worth noting that their label "universal" should not be interpreted as unrestricted transferability: systematic softening and failures for defects or vibration property have been documented.[59,60] The scalable and adaptive computational foundations are essential for enabling autonomous discovery loops that continuously integrate simulation, experiment, and scientific decision-making. However, achieving higher levels of autonomy may ultimately require new computational frameworks and scientific representations, as discussed in the following section.

### 4.4 New computational representations and scientific abstractions

Current computational materials agents operate largely within representations, governing equations, and codes designed by humans. Automating their use can deliver substantial value, but the highest proposed autonomy levels imply something qualitatively different: discovering new variables, governing relations, coarse-graining strategies, or algorithms that make previously inaccessible phenomena predictable. A historical analogue is Hilbert's sixth problem, which called for the mathematical foundation of physics through the connection between microscopic and macroscopic descriptions.

As we mentioned before, several emerging approaches provide bounded routes toward this capability. AI Feynman, for example, demonstrated that physics-inspired decomposition can recover compact symbolic relations from numerical observations,[61] while sparse identification methods have shown that governing equations of nonlinear dynamical systems can be inferred from data under appropriate structural assumptions.[62] Physics-informed machine learning provides another route by incorporating conservation laws, symmetries, differential equations, and other physical constraints into data-driven models, thereby narrowing the hypothesis space toward physically admissible representations.[63] Neural operators can augment, or even replace, existing numerical simulators in many applications, including material modelling, providing speedups of four to five orders of magnitude.[64] At the software level, by pairing a pretrained LLM with a systematic evaluator, FunSearch can discover a program that constructs a previously unknown mathematical object.[65]

These methods, however, still generally operate within problem formulations, hypothesis spaces, representations, or physical priors that are at least partly specified by humans. The more ambitious goal of L5 autonomy is to enable agents to identify when existing representations are inadequate and to formulate new descriptors, governing relations, computational abstractions, or solution strategies that expand the range of materials phenomena that can be predicted and understood.

**4.5 Benchmarking and continual improvement of autonomous agents**

Establishing standardized benchmarks for autonomous materials agents is essential for comparing capabilities across platforms and for defining meaningful progress toward higher autonomy. Existing AI benchmarks often emphasize task completion or output-level accuracy, whereas scientific agents must be evaluated across the workflow itself. Recent benchmarks have begun to assess multi-step and open-ended scientific tasks rather than question answering alone.[66,67] For computational materials agents, future benchmarks should therefore assess not only whether a calculation is completed correctly, but also whether the agent can formulate meaningful hypotheses, select appropriate computational strategies, operate heterogeneous tools, manage uncertainty and failures, adapt its workflow based on intermediate results, and, at the highest autonomy levels, extend its own computational capabilities.

Beyond evaluation, autonomous agents also require mechanisms for continual improvement through experience. Reinforcement learning (RL) provides a natural framework for such sequential decision-making, in which an agent interacts with an environment and learns a policy that maximizes cumulative reward. RL has already been explored in materials and chemical discovery for adaptive navigation of high-dimensional design spaces and sequential optimization.[68,69] In an autonomous computational workflow, rewards could incorporate not only scientific objectives but also computational cost, reliability, uncertainty reduction, and reproducibility, allowing the agent to learn more effective research strategies over repeated interactions. Benchmarks for higher-level autonomy should consequently evaluate not only static task performance but also learning efficiency: whether an agent can reduce unnecessary calculations, improve method and tool selection, recover more effectively from failures, and achieve better scientific outcomes as experience accumulates. Such longitudinal evaluation will be necessary to distinguish genuine continual improvement from repeated execution of fixed workflows

### 4.6 Outlooks

The ultimate goal of autonomous computational materials discovery is not to replace human scientists, but to transform the way scientific knowledge is generated. In this future paradigm, researchers will increasingly focus on defining fundamental scientific objectives, establishing meaningful constraints, and interpreting discoveries, while autonomous agents handle increasingly complex cycles of hypothesis generation, computation, validation, and optimization. By combining physical knowledge, computational capability, and autonomous reasoning, agentic AI may enable a new era of materials discovery in which humans and intelligent computational partners jointly explore previously inaccessible regions of materials space.

**Acknowledgements**

This work is financially supported by the Advanced Materials-National Science and Technology Major Project (2025ZD0618802), and National Natural Science Foundation of China (No.52332005).

**Author contributions**

Linggang Zhu and Zhimei Sun conceptualized, acquired funding for the project. Linggang Zhu wrote the first manuscript and prepared the figures. All authors contributed to reviewing and editing the manuscript.

**Competing interests**

The authors declare no conflict of interest.

## References

1. Louie, S. G., Chan, Y.-H., da Jornada, F. H., Li, Z. & Qiu, D. Y. Discovering and understanding materials through computation. *Nat. Mater.* **20**, 728-735 (2021).
2. Stephen, S. Computational materials design. *Nat. Mater.* **20**, 727-727 (2021).
3. Horton, M. K. *et al.* Accelerated data-driven materials science with the Materials Project. *Nat. Mater.* **24**, 1522-1532 (2025).
4. Ong, S. P. *et al.* Python Materials Genomics (pymatgen): A robust, open-source python library for materials analysis. *Comput. Mater. Sci.* **68**, 314-319 (2013).
5. Pizzi, G., Cepellotti, A., Sabatini, R., Marzari, N. & Kozinsky, B. AiiDA: automated interactive infrastructure and database for computational science. *Comput. Mater. Sci.* **111**, 218-230 (2016).
6. Jain, A. *et al.* Commentary: The materials project: A materials genome approach to accelerating materials innovation. *APL Materials* **1**, 011002 (2013).
7. Wang, G. *et al.* ALKEMIE: An intelligent computational platform for accelerating materials discovery and design. *Comput. Mater. Sci.* **186**, 110064 (2021).
8. Zhao, X.-G. *et al.* JAMIP: an artificial-intelligence aided data-driven infrastructure for computational materials informatics. *Sci. Bull.* **66**, 1973-1985 (2021).
9. Yang, X. *et al.* MatCloud: A high-throughput computational infrastructure for integrated management of materials simulation, data and resources. *Comput. Mater. Sci.* **146**, 319-333 (2018).
10. Szymanski, N. J. *et al.* An autonomous laboratory for the accelerated synthesis of inorganic materials. *Nature* **624**, 86-91 (2023).
11. Ramos, M. C., Collison, C. J. & White, A. D. A review of large language models and autonomous agents in chemistry. *Chemical Science* **16**, 2514-2572 (2025).
12. Dai, T. *et al.* Autonomous mobile robots for exploratory synthetic chemistry. *Nature* **635**, 890-897 (2024).
13. Ren, Z., Ren, Z., Zhang, Z., Buonassisi, T. & Li, J. Autonomous experiments using active learning and AI. *Nat. Rev. Mater.* **8**, 563–564 (2023).
14. Boiko, D. A., MacKnight, R., Kline, B. & Gomes, G. Autonomous chemical research with large language models. *Nature* **624**, 570-578 (2023).
15. Ament, S. *et al.* Autonomous materials synthesis via hierarchical active learning of

nonequilibrium phase diagrams. *Science Advances* **7**, eabg4930 (2021).

16 Montoya, J. H. *et al.* Autonomous intelligent agents for accelerated materials discovery. *Chemical Science* **11**, 8517-8532 (2020).

17 Gottweis, J. *et al.* Accelerating scientific discovery with Co-Scientist. *Nature* **655**, 487-496 (2026).

18 Li, C., Ran, N. & Liu, J. Agentic material science. *Journal of Materials Informatics* **6**, 10 (2026).

19 Gillon, R. Autonomy and the principle of respect for autonomy. *British medical journal (Clinical research ed.)* **290**, 1806-1808 (1985).

20 Vriza, A., Kornu, U., Koneru, A., Chan, H. & Sankaranarayanan, S. K. R. S. Multi-agentic AI framework for end-to-end atomistic simulations. *Digital Discovery* **5**, 440-452 (2026).

21 Ghafarollahi, A. & Buehler, M. J. Automating alloy design and discovery with physics-aware multimodal multiagent AI. *PNAS* **122**, e2414074122 (2025).

22 Ding, L., Carrillo, J.-M. & Do, C. ToPolyAgent: AI agents for coarse-grained bead-spring topological polymer simulations. *Digital Discovery* **5**, 901-909 (2026).

23 Alexander Timms , A. L., Antonis Antonopoulos, Antonis Mygiakis, Eleni Voulgari, Fearghal O'Donncha. Agentic AI for Digital Twin. *Proceedings of the AAAI Conference on Artificial Intelligence* **39**, 29703-29705 (2025).

24 Burr, C., Enzer, M., Shepherd, J. & Wagg, D. J. J. A. Agentic Digital Twins: A Taxonomy of Capabilities for Understanding Possible Futures. *Preprint at https://arxiv.org/abs/2601.18799* (2026).

25 Kalidindi, S. R., Buzzy, M., Boyce, B. L. & Dingreville, R. Digital Twins for Materials. *frontiers in Materials* **9**, 818535 (2022).

26 Yakutovich, A. V. *et al.* AiiDAlab – an ecosystem for developing, executing, and sharing scientific workflows. *Comput. Mater. Sci.* **188**, 110165 (2021).

27 Mathew, K. *et al.* Atomate: A high-level interface to generate, execute, and analyze computational materials science workflows. *Comput. Mater. Sci.* **139**, 140-152 (2017).

28 Curtarolo, S. *et al.* AFLOW: An automatic framework for high-throughput materials discovery. *Comput. Mater. Sci.* **58**, 218-226 (2012).

29 Liu, J. *et al.* VASPilot: MCP-facilitated multi-agent intelligence for autonomous VASP simulations. *Chin. Phys. B* **34**, 117106 (2025).

30 Yang, J., Kobayashi, Y. & Demura, M. AI agents for automating materials research: a case study of crystal plasticity simulations. *Science and Technology of Advanced Materials: Methods* **6**, 2630445 (2026).

31 Rothfarb, S. *et al.* Hierarchical Multi-agent Large Language Model Reasoning for Autonomous Heterogeneous Catalyst Discovery. *npj Comput. Mater.* **12**, 309 (2026).

32 Shuyi Jia, Chao Zhang & Fung, V. LLMatDesign: Autonomous Materials Discovery with Large Language Models. *Preprint at* https://arxiv.org/abs/2406.13163 (2024).

33 Lianhao Zhou *et al.* Toward Greater Autonomy in Materials Discovery Agents: Unifying Planning, Physics, and Scientists. *Preprint at* https://arxiv.org/abs/2506.05616 (2026).

34 Zhou, J., Xiao, B., Liu, Q., Liu, L. & Zhang, L. MatPC: Prompting Large Language Model, Crystal Structure Prediction, and First-Principles for Semantic-Driven Material Design. *ACS Appl. Mat. Interfaces* **17**, 44528-44540 (2025).

35 Pham, T. D. *et al.* Multi-Agent Orchestration for High-Throughput Materials Screening on a Leadership-Class System. *Preprint at* https://arxiv.org/abs/2604.07681 (2026).

36 Liang, H. *et al.* Real-time experiment-theory closed-loop interaction for autonomous materials science. *Science Advances* **11**, eadu7426 (2025).

37 Huang, H. *et al.* ALKEMIE Agent: an autonomous platform for computational materials design. *Preprint at* https://arxiv.org/abs/2608.15776 (2026).

38 Chaudhari, A., Ock, J. & Barati Farimani, A. Modular large language model agents for multi-task computational materials science. *Communications Materials* **7**, 131 (2026).

39 Wang, X. *et al.* Accelerating materials discovery via AI-Agent integration of large language models and simulation tools. *Journal of Materials Informatics* **6**, 9 (2026).

40 Zou, Y. *et al.* El Agente: An autonomous agent for quantum chemistry. *Matter* **8**, 102263 (2025).

41 Qian, J. *et al.* Digital Twin for Chemical Science: a case study on water interactions on the Ag(111) surface. *Nat. Comput. Sci.* **5**, 793-800 (2025).

42 Kalai, A. T., Nachum, O., Vempala, S. S. & Zhang, E. Evaluating large language models for accuracy incentivizes hallucinations. *Nature* **653**, 1047-1051 (2026).

43 Venugopal, V. & Olivetti, E. MatKG: An autonomously generated knowledge graph in Material Science. *Sci. Data* **11**, 217 (2024).

44 Mrdjenovich, D. *et al.* propnet: A Knowledge Graph for Materials Science. *Matter* **2**, 464-

480 (2020).

45 Marwitz, T. *et al.* Predicting new research directions in materials science using large language models and concept graphs. *Nature Machine Intelligence* **8**, 535-544 (2026).

46 Ghafarollahi, A. & Buehler, M. J. SciAgents: Automating Scientific Discovery Through Bioinspired Multi-Agent Intelligent Graph Reasoning. *Adv. Mater.* **37**, 2413523 (2025).

47 Oh, C. *et al.* Uncertainty Quantification in LLM Agents: Foundations, Emerging Challenges, and Opportunities. *Proceedings of the 64th Annual Meeting of the Association for Computational Linguistics* **1**, 16219-16250 (2026).

48 Kulichenko, M. *et al.* Uncertainty-driven dynamics for active learning of interatomic potentials. *Nat. Comput. Sci.* **3**, 230-239 (2023).

49 Lysogorskiy, Y., Bochkarev, A., Mrovec, M. & Drautz, R. Active learning strategies for atomic cluster expansion models. *Phys. Rev. Mater.* **7**, 043801 (2023).

50 Wellendorff, J. *et al.* Density functionals for surface science: Exchange-correlation model development with Bayesian error estimation. *Phys. Rev. B* **85**, 235149 (2012).

51 Parks, H. L., Kim, H.-Y., Viswanathan, V. & McGaughey, A. J. H. Uncertainty quantification in first-principles predictions of phonon properties and lattice thermal conductivity. *Phys. Rev. Mater.* **4**, 083805 (2020).

52 Otis, R. A. & Liu, Z.-K. High-Throughput Thermodynamic Modeling and Uncertainty Quantification for ICME. *JOM* **69**, 886–892 (2017).

53 Peherstorfer, B., Willcox, K. & Gunzburger, M. Survey of Multifidelity Methods in Uncertainty Propagation, Inference, and Optimization. *SIAM Rev.* **60**, 550-591 (2018).

54 Sabanza-Gil, V. *et al.* Best practices for multi-fidelity Bayesian optimization in materials and molecular research. *Nat. Comput. Sci.* **5**, 572-581 (2025).

55 Deng, B. *et al.* CHGNet as a pretrained universal neural network potential for charge-informed atomistic modelling. *Nature Machine Intelligence* **5**, 1031-1041 (2023).

56 Ilyes Batatia, D. P. K., Gregor Simm, Christoph Ortner, Gabor Csanyi. MACE: Higher Order Equivariant Message Passing Neural Networks for Fast and Accurate Force Fields. *Advances in Neural Information Processing Systems 35 (NeurIPS 2022)* (2022).

57 Li, T. *et al.* DPA4: Pushing the Accuracy-Cost Frontier of Interatomic Potentials with EMFA SO(2) Convolution. *Preprint at https://arxiv.org/abs/2606.02419* (2026).

58 Liang, T. *et al.* NEP89: universal neuroevolution potential for inorganic and organic materials across 89 elements. *Nat. Comput. Sci.* **6**, 789-801 (2026).

59 Deng, B. *et al.* Systematic softening in universal machine learning interatomic potentials. *npj Comput. Mater.* **11**, 9 (2025).

60 Yu, H., Giantomassi, M., Materzanini, G., Wang, J. & Rignanese, G.-M. Systematic assessment of various universal machine-learning interatomic potentials. *Materials Genome Engineering Advances* **2**, e58 (2024).

61 Udrescu, S.-M. & Tegmark, M. AI Feynman: A physics-inspired method for symbolic regression. *Science Advances* **6**, eaay2631 (2020).

62 Brunton, S. L., Proctor, J. L. & Kutz, J. N. Discovering governing equations from data by sparse identification of nonlinear dynamical systems. *PNAS* **113**, 3932-3937 (2016).

63 Karniadakis, G. E. *et al.* Physics-informed machine learning. *Nature Reviews Physics* **3**, 422-440 (2021).

64 Azizzadenesheli, K. *et al.* Neural operators for accelerating scientific simulations and design. *Nature Reviews Physics* **6**, 320-328 (2024).

65 Romera-Paredes, B. *et al.* Mathematical discoveries from program search with large language models. *Nature* **625**, 468-475 (2024).

66 Ziru Chen, S. C., Yuting Ning, Qianheng Zhang, Boshi Wang, Botao Yu, Yifei Li, Zeyi Liao, Chen Wei, Zitong Lu, Vishal Dey, Mingyi Xue, Frazier N. Baker, Benjamin Burns, Daniel Adu-Ampratwum, Xuhui Huang, Xia Ning, Song Gao, Yu Su, Huan Sun. ScienceAgentBench: Toward Rigorous Assessment of Language Agents for Data-Driven Scientific Discovery. *International Conference on Learning Representations 2025 (ICLR 2025)* (2025).

67 Liu, T. *et al.* Benchmarking AI Agents for Addressing Scientific Challenges Across Scales. *Preprint at* https://arxiv.org/abs/2606.12736 (2026).

68 Volk, A. A. *et al.* AlphaFlow: autonomous discovery and optimization of multi-step chemistry using a self-driven fluidic lab guided by reinforcement learning. *Nat. Commun.* **14**, 1403 (2023).

69 Xian, Y. *et al.* Unlocking the black box beyond Bayesian global optimization for materials design using reinforcement learning. *npj Comput. Mater.* **11**, 143 (2025).